\documentclass[aps,prb,twocolumn,showpacs]{revtex4-1}

\usepackage{natbib}
\usepackage{amssymb}
\usepackage{amsmath}
\usepackage{graphicx}
\usepackage{dcolumn}
\usepackage{bm}
\usepackage{color}

\begin{document}

\title{Establishing characteristic behavior of voltage control of magnetic anisotropy by ionic migration}
\author{F.~Ibrahim}

\author{A.~Hallal}

\author{B.~Dieny}

\author{M.~Chshiev}
\affiliation{Univ. Grenoble Alpes, CEA, CNRS, Grenoble INP, INAC-Spintec, 38000 Grenoble, France}

\begin{abstract}
A characteristic dependence of voltage control of perpendicular magnetic anisotropy (VCMA) on oxygen migration at Fe/MgO interfaces was revealed by performing systematic {\it ab initio} study of the energetics of the oxygen path around the interface. We find that the surface anisotropy energy exhibits a Boltzmann sigmoidal behavior as a function of the migrated O-atoms concentration. The obtained variation of the VCMA efficiency factor $\beta$ reveals a saturation limit beyond a critical concentration of migrated O, about $54\%$, at which the anisotropy switches from perpendicular to in plane. Furthermore, depending on the range of variation of the applied voltage, two regimes associated with reversible or irreversible ions displacement are predicted to occur, yielding different VCMA response. 
According to our findings, one can distinguish from the order of magnitude of $\beta$ the VCMA driving mechanism: an effect of several tens of fJ/(V.m) is likely associated to charge-mediated effect combined with slight reversible oxygen displacements whereas an effect of the order of thousands of fJ/(V.m) is more likely associated with irreversible oxygen ionic migration.
\end{abstract}
\maketitle

\section{Introduction}
Magnetization switching using spin-polarized currents via the spin transfer torque (STT) effect has
achieved remarkable progress ~\cite{Slonczewski,Berger,Myers}. However, the energy required to write
in STT-magnetic random access memories (MRAM) is still rather large (of the order of 100~fJ per write 
event) compared to typical write energy of volatile memories in complementary metal oxide semiconductor (CMOS) technology. An alternative strategy for manipulating magnetization with low power consumption relies on applying electric fields (E-field) rather than currents. Several experimental reports have demonstrated electric-field control of magnetic properties, among which 
those evidencing control of the perpendicular magnetic anisotropy (PMA). The latter originates from the spin-orbit interaction and electronic hybridization between oxygen and the magnetic transition metal orbitals across the interface~\cite{Yang,Dieny}. The voltage control of PMA (VCMA) is of particular importance to realize fast and low-power-consumption magnetization switching~\cite{Weisheit, Endo, Shiota, Amiri, Meng, Kita, Maruyama,Nozaki,Wang,Rajanikanth}. In particular, a strong impact of the electric field on the interfacial PMA in Fe(Co)/MgO-based systems was reported~\cite{Endo, Shiota, Amiri, Meng, Kita, Maruyama,Nozaki,Wang,Rajanikanth}. Meanwhile, theoretical studies have addressed the origin of this effect which was attributed to the spin-dependent 
screening of the electric field in ferromagnetic metal films~\cite{Duan}, and to the change in the relative occupancy
of the 3d-orbitals of Fe atoms associated to the electrons accumulation or depletion 
at the Fe/MgO interface~\cite{Niranjan,Nakamura1}. Furthermore, the effect was shown 
to be correlated with the existence of a spontaneous interfacial electric dipole~\cite{Ibrahim}. 
Typical calculated values for the charge-mediated PMA variation under electric field characterized 
by the parameter $\beta$ are of the order of tens of fJ/(V.m)~\cite{Ibrahim} which agrees with the 
experimental observations of references~\cite{Endo, Shiota, Amiri, Meng, Kita, Maruyama,Nozaki,Wang,Rajanikanth}. 

In this paper, we present a first-principles study of the tuning of the interfacial PMA by oxygen (O)
migration across the Fe/MgO interface. After describing the method in Section~\ref{Methods}, the charge-mediated VCMA effect
was calculated for both over-oxidized and oxygen-migrated Fe/MgO interfaces in Section~\ref{Chargemediated}. We confirm that the underlying microscopic mechanisms are sensitive to the oxidation conditions at the interface. Yet, the strength of the effect is found to be of the same order of magnitude for both types of interfaces (~a few tens of 
fJ/(V.m)). In Section~\ref{Omigrated}, we studied the variation of PMA associated with oxygen migration across the Fe/MgO interface. For that, the energetics of the O migration path across the interface was investigated and the 
impact of the O position on the PMA calculated. It is found that the PMA value, as well as its on-site resolved contributions, are highly affected by the O migration. 
Besides, we show in Section~\ref{Concentrationdep} that the PMA variation induced by O migration shows a Boltzmann sigmoidal behavior as a function of the concentration of migrated O. 
Interestingly, depending on the amplitude of the applied voltage variation, two regimes could be distinguished associated with reversible or irreversible oxygen ions displacement yielding different voltage controlled PMA response.   In the irreversible case, O-migration mediated VCMA can reach thousands of fJ/(V.m) consistent with the experimental observations \cite{Bauer}.           


\begin{figure*}[ht]
  \centering
     \includegraphics[width=0.9\textwidth]{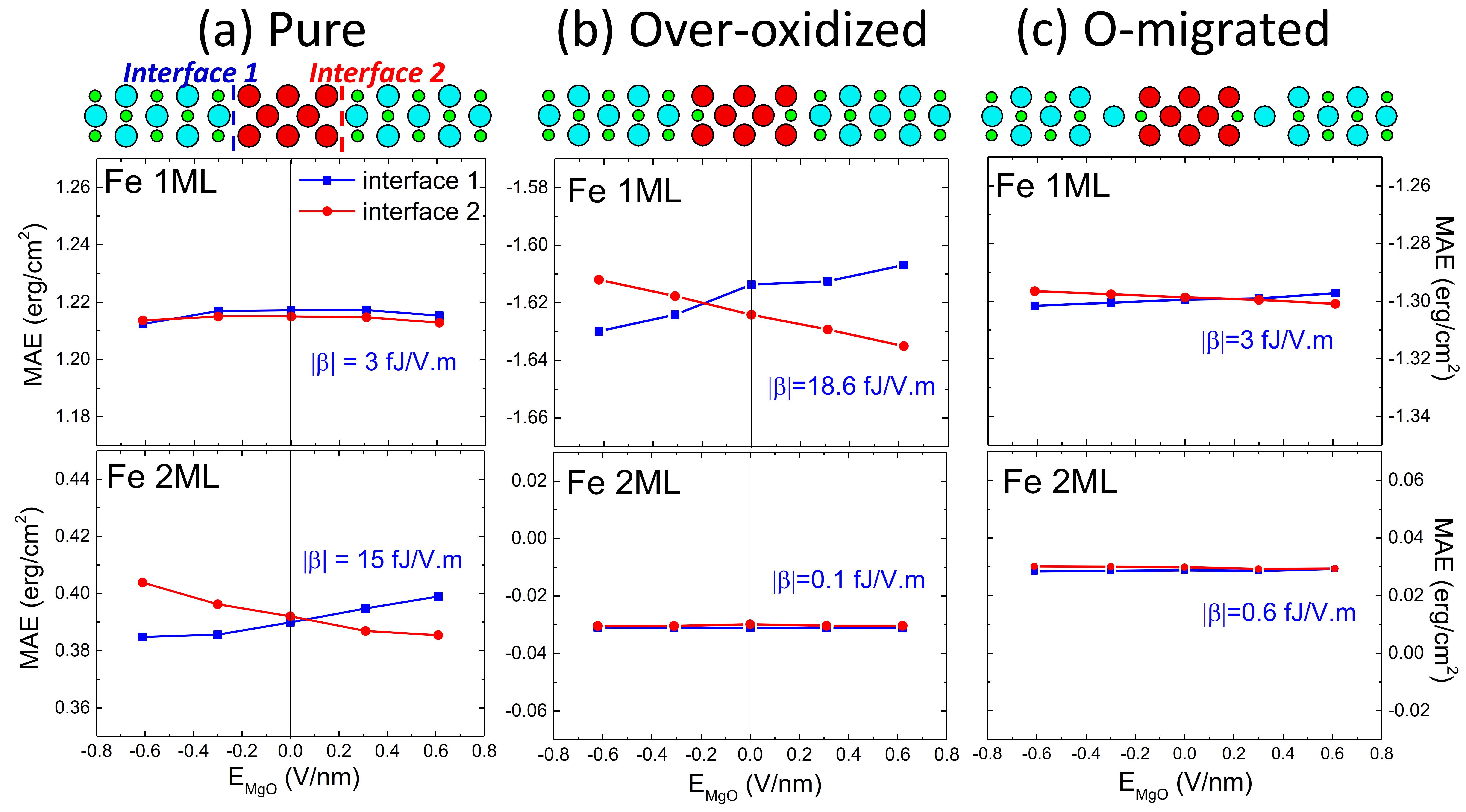}
  \caption{(Color online) Layer-resolved variation of MAE of both interfaces as a function of electric field 
  in MgO, calculated for the first and second Fe MLs in pure (a), over-oxidized (b), and O-migrated (c) 
  MgO/Fe/MgO sandwich. The upper panels show the supercell structure used in each case (Red=Fe, green=O, 
  Light blue=Mg). The slope ($\beta$) of the MAE variation with electric field is calculated and displayed 
  for each ML. Blue(red) indicates interface 1(2).}
  \label{fig1}
\end{figure*}  

\section{Methods}\label{Methods}
Our first-principles calculations are based on the projector-augmented wave (PAW) method \cite{Blochl} as implemented in the VASP package \cite{Kresse1,Kresse2,Kresse3} using the generalized gradient approximation \cite{Perdew} and including spin-orbit coupling. A kinetic energy cutoff of 500 eV has been used for the plane-wave basis set and a $25\times 25\times 1$ K-point mesh to sample the first Brillouin zone. The  electric field, applied perpendicular to the supercell, is introduced as a dipole layer placed in the 
vacuum region of the supercell as proposed by the dipole layer method \cite{Neugebauer} and varied 
between $-2$ V/nm and $2$ V/nm. The supercell comprises 5 Fe monolayers (ML) sandwiched between 5ML 
of MgO followed by a vacuum layer. This structure provides the opportunity to compare the physical 
properties of two different Fe/MgO interfaces simultaneously in one calculation as interpreted 
in Ref.~\cite{Ibrahim}. The over-oxidized interface is modeled by inserting an additional O atom at
the interfacial Fe monolayer. In contrast, the oxygen-migrated interface is described by moving an
O atom from the interfacial MgO plane towards the Fe layer (Fig.~\ref{fig1}). The in-plane lattice 
constant was fixed to that of Fe (i.e. $a=2.87$~\AA), while the structure was relaxed in the absence
of electric field until the forces became smaller than $1$ meV/\AA. The orbital and layer-resolved 
magnetic anisotropy contributions are evaluated following \cite{Hallal2,Hallal1}. The number of k-points is adjusted so that the magnetic anisotropy energy is calculated with an accuracy of $\pm0.002$ erg/cm$^{2}$. For the larger supercell used in Section~\ref{Concentrationdep} in order to investigate the effect of concentration of O migrating ions on the magnetic anisotropy, we used a $15\times15\times1$ K-point mesh which was able to reproduce the same values of the total and projected magnetic anisotropy energies as in the smaller ($1\times1$) supercell together with an accuracy of $\pm0.002$ erg/cm$^{2}$ in the calculated values.

\section{Charge-mediated VCMA}\label{Chargemediated}
The interfacial oxidation conditions have a strong impact on the PMA at Fe/MgO 
interfaces~\cite{Hallal1,Yang,Monso,Manchon,Rodmacq, Rodmacq2,Nistor} and it was pointed 
out that at over-oxidized Fe/MgO interfaces, the PMA may be altered by the electric 
field~\cite{Nakamura2}. However, a detailed description of the microscopic mechanisms of the 
electric field control of the PMA under different oxidation conditions is still lacking. In 
this context, we compare in Fig.~\ref{fig1} the variation of the layer-resolved contributions 
to magnetic anisotropy  energy (MAE) per interface as a function of the electric field in MgO 
for pure, over-oxidized and O-migrated Fe/MgO interfaces, shown respectively in Fig.~\ref{fig1}(a),(b) and (c). 
Strikingly, the sum of the contributions to the E-field induced MAE variation from the first and second Fe MLs 
is almost the same in the pure and over-oxidized interface (pure case:$\beta=3+15=18$~fJ/(V.m) vs over-oxydized 
case:$\beta=18.6+0.1=18.7$~fJ/(V.m)). However at the microscopic level, the situations are quite different. For pure interface, the second ML is mostly responsible for the E-field induced variation of MAE whereas for the over-oxidized interface, the first ML clearly plays the dominant role (cf. Fig.~\ref{fig1}(a) and (b)). Furthermore, for O-migrated interface, the E-field induced variation of MAE   is strongly reduced ($\beta=3+0.6=3.6$~fJ/(V.m))(Fig. \ref{fig1}(c)).

\begin{figure}[ht]
  \centering
     \includegraphics[width=0.45\textwidth]{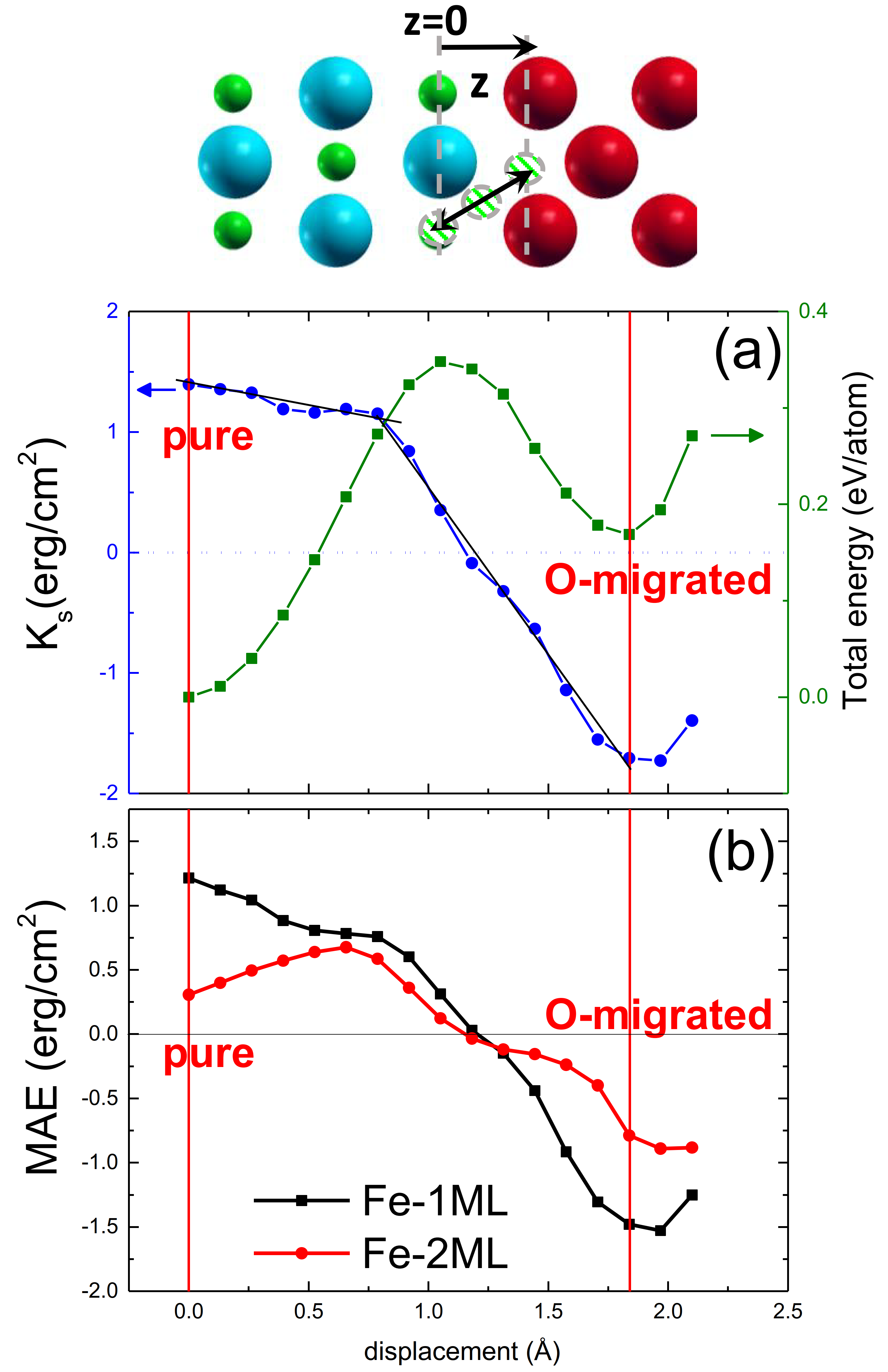}
  \caption{(Color online) Top: Zoom in "interface 1" of the MgO/Fe/MgO supercell illustrating the path of the O atom migrating across the Fe/MgO interface such that $z$ corresponds to the displacement of the O atom from the MgO plane.  (a) Calculated surface anisotropy $K_{s}$ 
  (blue) and relative variation of the total energy per atom (green)
  as a function of O atom displacement from the MgO plane toward the
  Fe interface. (b) Layer resolved magnetic anisotropy as a function
  of the O displacement across the Fe/MgO interface.}
  \label{fig2}
\end{figure}

\section{O-migration mediated VCMA}\label{Omigrated}
 
The charge-mediated VCMA at transition metal/oxide interfaces cannot explain
the large variation of MAE versus electric field reaching thousands of fJ/(V.m) as reported
recently~\cite{Rajanikanth,Bauer}. Such enhanced VCMA values can be attributed to a different mechanism associated
with E-field induced O-displacement around the transition metal/oxide interface and its impact on the interfacial
MAE~\cite{Dieny}. In particular, this mechanism can be supported by the fact that electric field extends almost over one interatomic distance from the insulator towards metal as it has been shown in case of Fe/MgO interfaces~\cite{Ibrahim}. Here we describe theoretically the ionic migration induced VCMA. For that, an oxygen migration path is created by moving the O atom starting from the initial 
pure interface [Fig.~\ref{fig1}(a)] towards the first Fe monolayer [Fig.~\ref{fig1}(c)]. At each 
point along this path, the total energy per atom relative to the initial state is calculated. The 
result is plotted by line-squares in Fig.~\ref{fig2} as a function of the O atom displacement 
from the MgO plane. Interestingly, the total energy shows a local minimum at an oxygen displacement 
of about $1.84$~\AA{} separated from the initial position by an energy maximum for a displacement z$_{c}$. 
The corresponding energy barrier that should be overcome to pass from pure interface to the O-migrated
one is about $0.35$~eV/atom. This value is likely overestimated since we consider here the motion of an
O atom in a quite narrow supercell. For larger supercell, a significant reduction of the energy barrier is expected ~\cite{Hualde} and will be discussed further. Besides, we assumed that the O migration path follows
a straight line between its initial and final position whereas the lowest energy path may be more complex
yielding a lower energy barrier. Nevertheless, despite the likely overestimation of our calculated energy
barrier height for O migration, the following discussion remains semi-quantitatively valid. Using our 
calculated value of the energy barrier for O migration across the Fe/MgO, we can estimate the force 
acting on the oxygen ion (charge 2e) and the corresponding critical electric field $E_c$ needed to 
overcome the energy barrier. This value is $E_c = \frac {\Delta V}{\Delta z}=1.9$~V/nm.
In Fig.~\ref{fig2}, the variation of $K_{s}$ versus O-displacement is also plotted. Two parts 
can be seen in this variation: a slight almost linear negative slope between the origin and O-displacement
of $0.8$~\AA{}, and a steep decrease afterwards till the O-migrated interface state is reached. 
Actually, two regimes of oxygen displacement can be distinguished.
i)If the electrical field is lower than $E_c$, the oxygen ion is reversibly moving around its equilibrium
position due to the electrostatic force exerted by the applied electrical field. As seen in Fig.~\ref{fig2},
the interfacial anisotropy depends on the exact position of the oxygen ion. Using O-displacement position $z_c$
corresponding to the barrier maximum, this yields a first $\beta$ value of 
($\beta\sim(dK_{s}/dz)*z_c/E_c\simeq 230$~fJ/(V.m)) which is already rather large and 
comparable to some experimentally obtained values. This regime being reversible, 
the associated time scale can be extremely short (approaching inverse phonon frequency 
i.e. THz regime). 
It is important to note that this effect has the same  influence as 
that due to charge accumulation/depletion. Indeed in the latter mechanism, an interfacial 
electron accumulation in the Fe layer caused by an electrical field pointing out from the MgO 
layer towards the Fe layer causes a decrease in the interfacial anisotropy. For the O-displacement mechanism, such a field pulls the oxygen further away from the 
interface which reduces the interfacial anisotropy.
The significant variations in the 
experimentally measured $\beta$ values reported by various groups may at least partly be 
ascribed by variations in the relative influence of these two competing mechanisms.
ii) A second regime is then expected when the applied electrical field is large enough to
pull the oxygen atom above the migration barrier so that it relaxes towards its new position
within the interfacial Fe plane (position corresponding to the O-migrated interface). In this
case, using the aforementioned $E_c$ value and the variation in $K_{s}$ between the cases of 
pure and O-migrated interfaces, we estimate $\beta=-1600\pm50$~fJ/(V.m) which is in good agreement
with the experimental value in Ref.~\cite{Rajanikanth} ($\beta=-1150$~fJ/(V.m)) and of the same order 
of magnitude as in Ref.~\cite{Bauer}.
Although the experimental data of Ref.~\cite{Bauer} are obtained for Co/GdOx interface which is quite different from the system we are investigating, the origin of magnetic anisotropy in Co/Ox or Fe/Ox is the same, namely Oxygen ion orbitals hybridization with those of the transition metal. It is shown theoretically and experimentally that over-oxidation in both ferromagnetic films leads to a strong reduction of PMA Ref.~\cite{Hallal1, Yang}.  In asymmetric structure, application of electric field in one direction will result in a VCMA driven by ionic migration in addition to the charge mediated effect. However, application of electric field in the opposite direction will result in a VCMA driven by charge mediated effect only since migration of oxygen atom into the MgO barrier will have small influence on the PMA compared to migration to the iron interface Ref~\cite{Hallal1}. This asymmetry could explain the observed nonlinearity of VCMA reported for V/Fe/MgO structure in Ref.~\cite{Rajanikanth}.  

One can point out that the estimated barrier height for oxygen ion migration is about $0.35$~eV which is more than an order of magnitude larger than room temperature energy. In other words, the energy of the electric field is much larger than $K_{B}T$. In such high regime of electric fields compared to common diffusion studies, our system cannot be considered at thermodynamic equilibrium. This justifies neglecting temperature effects in our present approach. Indeed, in VCMA experiments on magnetic tunnel junctions, the applied electric fields are of the order of $0.1$~V/nm, i.e. one tenth of the critical electric field yielding dielectric breakdown in oxides ($~10^{9}$~V/m). Such high electric field can have a strong influence on oxygen mobility since it is so close to the dielectric breakdown. In contrast to that, thermal fluctuations yielding crystal vibrations at room temperature are weak. Since we are still much below the oxide melting temperature, thermal activation is not the main mechanism driving atomic mobility in our case. Nevertheless, it can assist the electric field and help to overcome the barrier. In this case, because the oxygen displacement is irreversible, the associated time-scale of the anisotropy variation can be much longer as observed in the experiments~\cite{Bauer}. In Ref.~\cite{Bauer} the authors demonstrate how a small change in temperature and gate voltage can improve device response times by orders of magnitude. However, temperature alone cannot irreversibly change the magnetic properties of the interface. Thus annealing could shorten the response time significantly but at the end, thermal activation is not the main driving force behind the irreversible ionic motion. Application of electric field is therefore necessary to overcome the barrier. Furthermore, experiments on FeCo/MgO show that it is possible to change the oxidation state of the interface by electric field modulation without thermal activation~\cite{Bonell}. 
We should also note that although assuming a straight path of the O atom across the Fe/MgO interface is likely oversimplified, we believe that it is a good approximation which helps understanding the physics ruling the VCMA effect. For instance, if there exists another path with a lower energy barrier, then the VCMA rate would be even larger than our estimated value.

\begin{figure}[ht]
  \centering
     \includegraphics[width=0.45\textwidth]{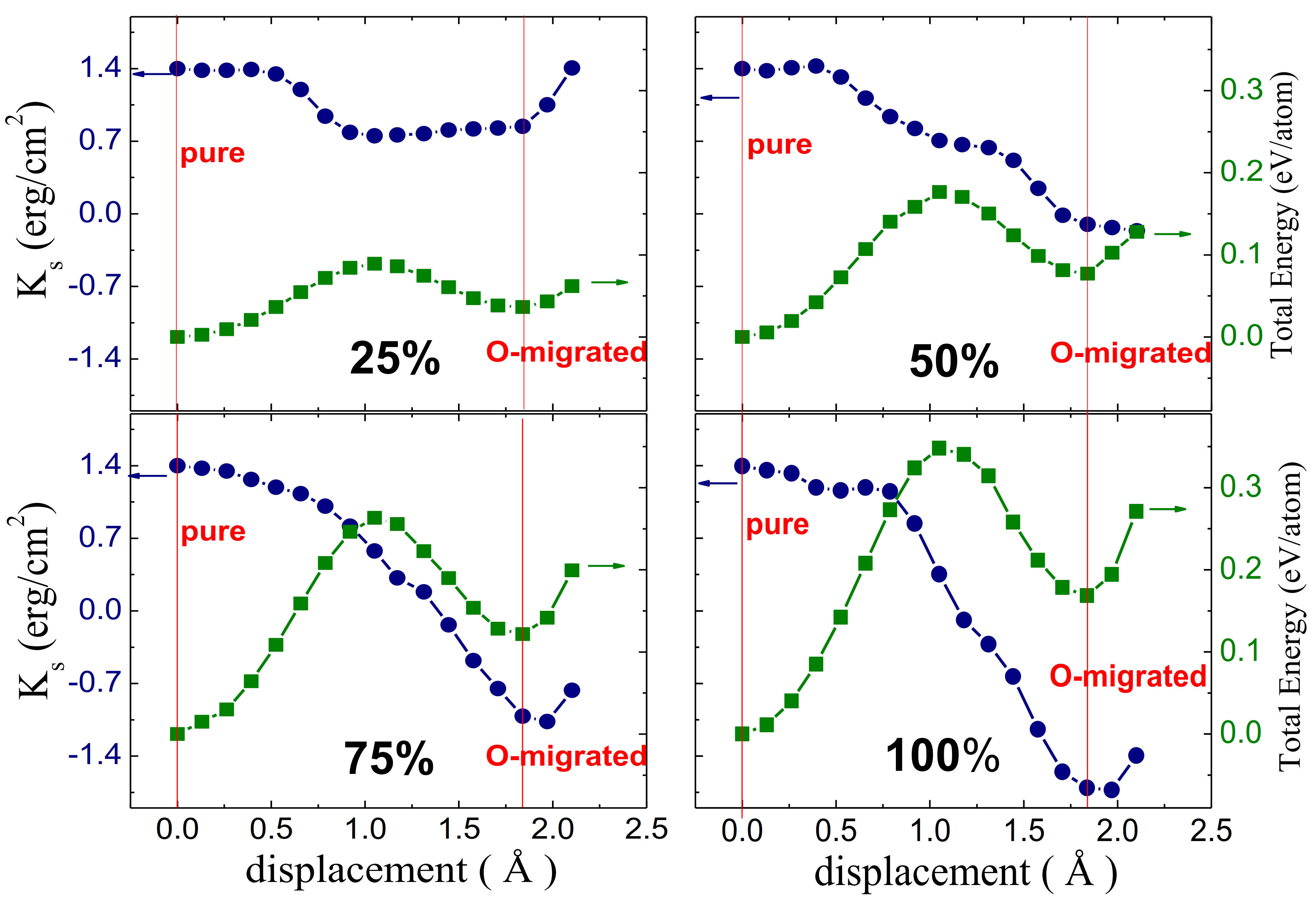}
  \caption{(Color online) Calculated surface anisotropy $K_{s}$ (blue) and relative variation 
  of the total energy per atom (green) as a function of O atom displacement from the MgO plane toward 
  the Fe interface for different concentration of O-migrating atoms. }
  \label{fig3}
\end{figure}

In Fig.~\ref{fig2}(b) we plot the layer-resolved MAE contributions into $K_{s}$ as a function 
of the O-displacement shown for the 1st and 2nd Fe ML. It can be seen that the slight negative
slope region of the $K_{s}$ variation (displacement$<0.8A$) originates from a partial balance 
between the decrease and the counter increase of the MAE respectively in 1st and 2nd Fe ML. 
On the contrary, the simultaneous decrease of the MAE of both layers observed for larger oxygen
displacement results in the steep decreases of $K_{s}$.


\section{Influence of concentration of O-migrated ions on VCMA}\label{Concentrationdep}

\begin{figure}[ht]
  \centering
     \includegraphics[width=0.45\textwidth]{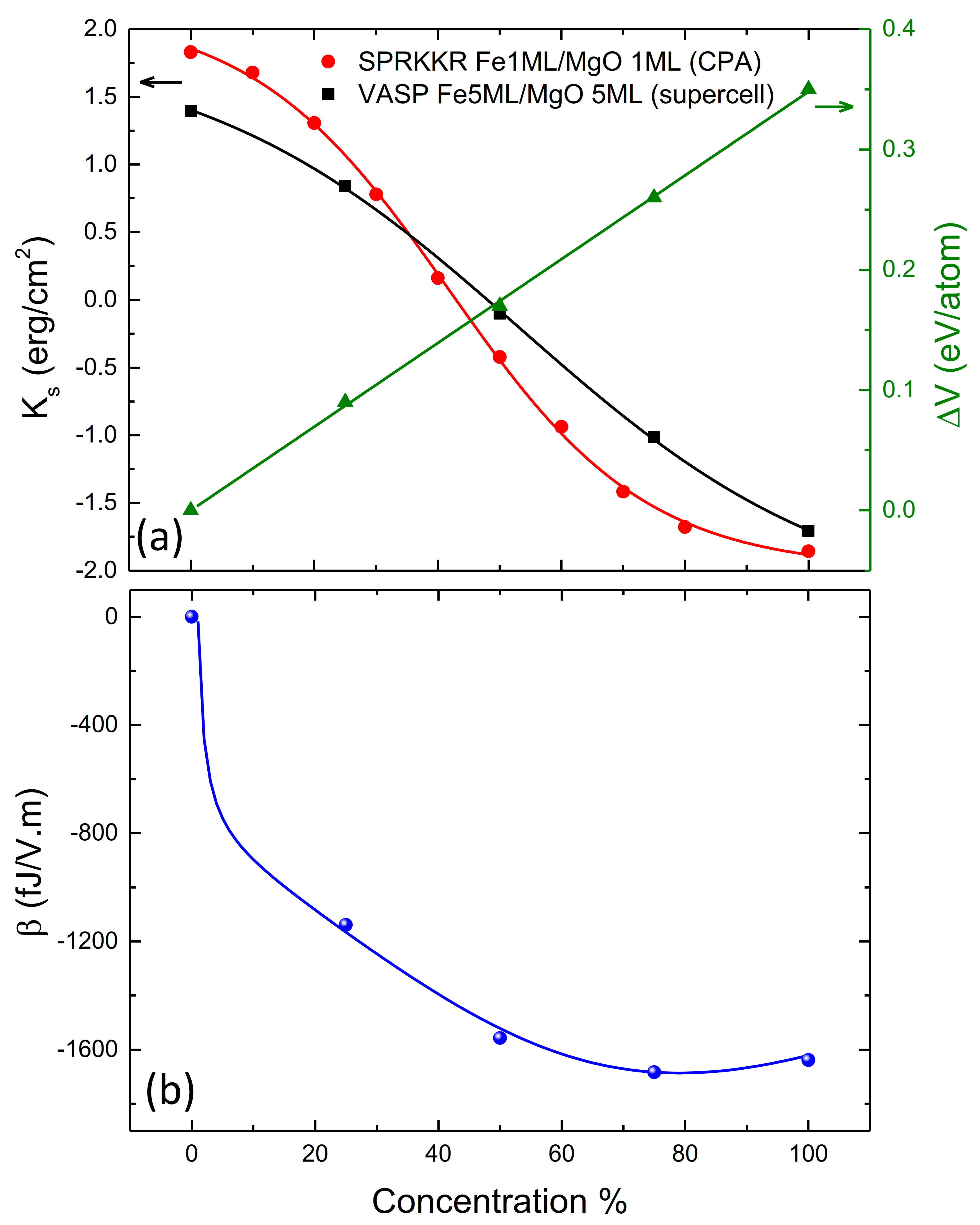}
  \caption{(Color online) (a) Energy barrier $\Delta V$ (triangles) and magneic anisotropy $K_s$ calculated by VASP (squares) and SPRKKR (circles); (b) VCMA coefficient $\beta$ due to oxygen ionic-migration versus the concentration of migrated oxygen at Fe/MgO interface.}
  \label{fig4}
\end{figure}  

To assess the validity of the aforementioned discussion, we now consider a larger supercell, $2\times2$ MgO(5ML)/Fe(5ML)/MgO (5ML).
This allows to investigate the effect of concentration of O migrating ions on the magnetic anisotropy as well as on the VCMA amplitude.
The total energy per atom 
relative to the initial state and the surface anisotropy is calculated 
for different O-migrating ion concentration and plotted in Fig.~\ref{fig3} as a function of the O displacement from the MgO plane. Similar to the previous results for smaller supercell,
the total energy shows a local minimum at an O displacement of about
$1.84$~\AA{} for all the calculated concentrations. As expected, the height of the energy barrier decreases with the decrease of the concentration
of O-migrating atoms such that the critical electric field $E_c$ corresponding to $25\%$ is about $0.5$~V/nm. As in the smaller supercell, the same discussion of the presence of two regimes of oxygen displacement (reversible and irreversible) applies for different O-migrating concentrations. In the reversible regime, the impact of O displacement on $K_s$ is evident for
large concentrations of migrating O atoms, above $50\%$, whereas for lower concentrations, displacing O atoms by a few tens of pico-meters will not affect the PMA of Fe at MgO interface as seen in Fig~\ref{fig3}.

For irreversible processes, the variation of both the energy barrier $\Delta V$ and $K_{s}$ as a function of the migrated O concentration are plotted in Fig~\ref{fig4}(a). $\Delta V$ (green triangles) is perfectly fitted by an increasing linear function (solid line) such that 

\begin{equation}
 \Delta V= \alpha c
\label{eqn1}
\end{equation}

 where $c$ stands for the migrated O concentration. However, $K_{s}$ (black squares) shows a behavior which best fits to a Boltzmann sigmoidal function (solid black line) of the form: 

\begin{equation}
 K_{s}= \frac{K_{0}-K_{f}}{1+e^{\frac{(c-c_{0})}{\omega}}}+K_{f}
\label{eqn2}
\end{equation}

\begin{figure}[h]
  \centering
     \includegraphics[width=0.45\textwidth]{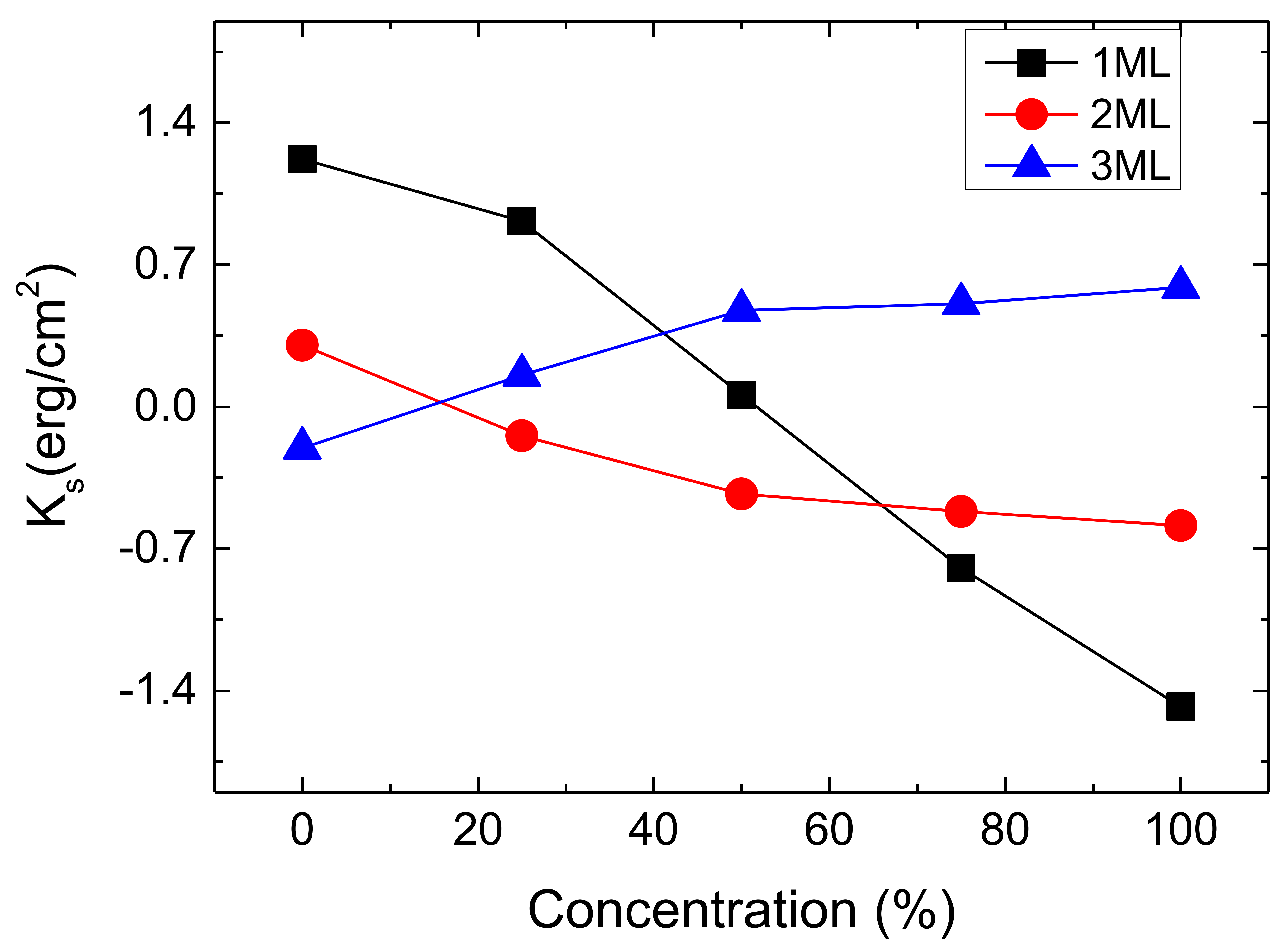}
  \caption{(Color online) Layer resolved magnetic anisotropy $K_s$ as a function of the concentration of migrated oxygen at Fe/MgO interface.}
  \label{fig5}
\end{figure}

In this function, $K_{0}$ and $K_{f}$ correspond to the surface anisotropy values of the pure and O-migrated interfaces, respectively. Correspondingly, $c_0$ and $\omega$ represent the inflection concentration point and the steepness. Interestingly, we find a value of $c_{0}\backsimeq 54\%$ which corresponds to the critical concentration of migrated O beyond which the anisotropy switches from perpendicular to in plane. 
We further verified the obtained dependency of the surface anisotropy on the migrated O concentration by performing calculations based on the coherent potential approximation (CPA) conveniently implemented within the SPRKKR package~\cite{SPR2,SPR3,SPR4}. We used a supercell comprised of Fe monolayer on top of MgO monolayer, and found a similar trend of $K_{s}$ behavior as a function of migrated O concentration (red curve in Fig~\ref{fig4}(a)) supporting our aforementioned findings. The choice of a monolayer is justified since the layer resolved contributions of $K_s$ as a function of migrated O concentration reveal that the total $K_s$ is mainly contributed by the first Fe monolayer trend as function of the migrated O concentration, while the second and third Fe monolayers yield compensated changes as shown in Fig.~\ref{fig5}.

Indeed, the ionic migration driven VCMA factor $\beta$ can be estimated for each value of migrated O concentration as $\beta=\frac{\Delta K_{s}}{\Delta V}\Delta z$ and shown in Fig.~\ref{fig4}(b) by blue circles. The calculated values of $\beta$ are fitted to the ratio of the functions in Eq.~\ref{eqn1} and Eq.~\ref{eqn2} to obtain the dependency of the VCMA on the migrated O concentration as:

\begin{equation}
 \beta = \beta_{max}\frac{(1+e^{\frac{-(c-c_{0})}{\omega}})^{-1}}{c}
 \label{eqn3}
\end{equation}

where $\beta_{max}=\frac{\Delta K_{max}}{\alpha} \Delta z \simeq -1600 \pm 50$~fJ/(V.m). This value represents the saturation of the ionic-migration driven VCMA effect which is attained once the PMA switches to in plane direction.  This relation between the VCMA and the concentration of O migrated ions in Fe/MgO brings about a new perspective of the VCMA effect. 

\begin{figure}[h]
  \centering
     \includegraphics[width=0.45\textwidth]{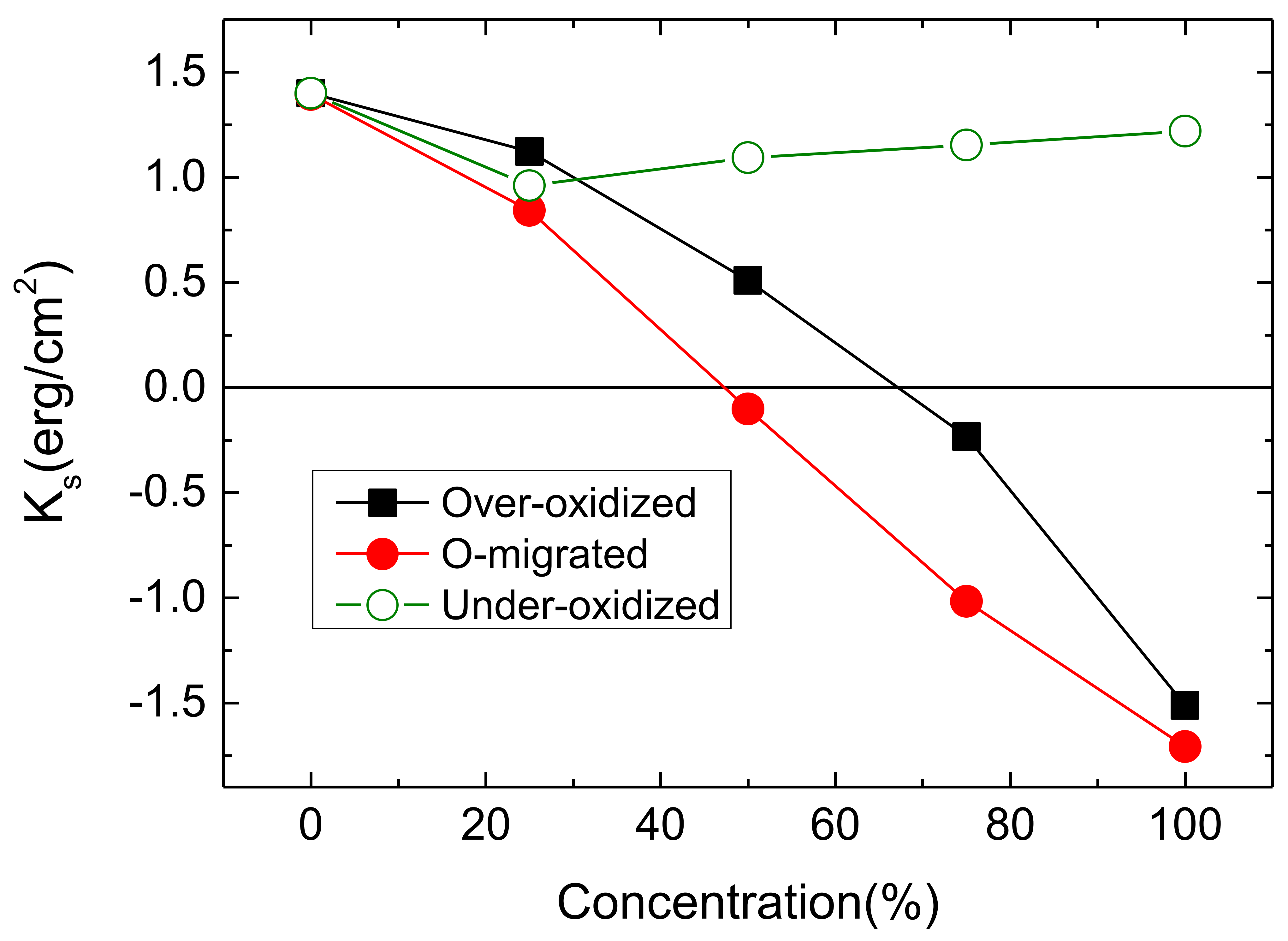}
  \caption{(Color online) Variation of the magnetic anisotropy $K_s$ as a function of concentration of over-oxidation (squares), under-oxidation (open circles), and migrated oxygen (closed circles) at Fe/MgO interface.}
  \label{fig6}
\end{figure}

To further understand the sigmoidal trend of $K_s$ as a function of the O migration concentration, we performed additional calculations of $K_s$ as a function of O concentration at over- and under-oxidized Fe/MgO interfaces. The results in Fig.~\ref{fig6} reveal a parabolic behavior of the over-oxidized case and a non-monotonous behavior of the under-oxidized one. In fact, the O migration process can be viewed as a collective effect of an over- and under-oxidation occurring simultaneously at a Fe/MgO interface. 

\section{Conclusion}\label{Conclusion}
In summary, we have presented a detailed study of the mechanisms underlying the electric-field impact on the PMA at Fe/MgO interfaces depending on interfacial oxidation conditions. The charge-mediated effect is found to be substantially weak in all considered cases. We have demonstrated that the O-migration across the Fe/MgO interface can provide a much more effective way to tune the PMA. Inspired by recent experiments, we propose the electric field as a possible driving force for O-migration and support this argument from the energetics point of view. Two regimes are expected: a reversible one under moderate electrical field and an irreversible one under larger electrical field. 
Besides, the VCMA rate is influenced by the concentration of migrated O ions in a characteristic way revealing a saturation limit beyond a critical concentration of O-migration, about $54\%$, at which the anisotropy switches from perpendicular to in plane. Interestingly, the estimated VCMA rate associated with interfacial O migration exceeds thousand of fJ/(V.m) in the regime of irreversible O-migration in agreement with experimental reports. Following those results, one can distinguish from the order of magnitude of $\beta$ which mechanism is driving the VCMA: An effect of several tens of fJ/(V.m) is likely associated with charge-mediated effect combined with slightly reversible oxygen displacement whereas an effect of the order of thousands of fJ/(V.m) is more likely associated with irreversible oxygen ionic migration effect.

This work has been supported by the ANR Project ELECSPIN (ANR-16-CE24-0018) and partly by the ERC Advanced Grant Project MAGICAL No. 669204.

\end{document}